# Quantum-Fluctuation-Initiated Coherence in Multi-Octave Raman Optical Frequency Combs

Y. Y. Wang, Chunbai Wu[1], F. Couny, M. G. Raymer[1] and F. Benabid*

*Gas-Phase Photonic Materials Group, Centre for Photonics and Photonic Materials, University of Bath, Claverton Down, Bath, BA2 7AY, UK*

[1] *Department of Physics and Oregon Center for Optics, University of Oregon, Eugene, Oregon 97403 US*

We show experimentally and theoretically that the spectral components of a multi-octave frequency comb spontaneously created by stimulated Raman scattering in a hydrogen-filled hollow-core photonic crystal fiber exhibit strong self coherence and mutual coherence within each 12 ns driving laser pulse. This coherence arises in spite of the field's initiation being from quantum zero-point fluctuations, which causes each spectral component to show large phase and energy fluctuations. This points to the possibility of an optical frequency comb with nonclassical correlations between all comb lines.

Optical frequency combs play important roles in classical and quantum optics. In the classical case they enable the synthesis of ultrastable frequency references [1] and ultrashort optical pulses, the latter of which entails the production of a multi-octave optical comb-like spectrum of mutually coherent lines [2]. This is based either on high harmonic generation in noble gases [3] or high-order stimulated Raman scattering (SRS) from hydrogen [4, 5]. In both processes, the medium coherence is externally driven so quantum vacuum fluctuations play no direct role in the comb generation.

SRS provides a unique juncture to explore the boundary between attosecond science and quantum optics. In SRS a microscopic quantum triggering event could be amplified to the macroscopic level. Early work on the statistical properties of SRS showed that when a spontaneously emitted Stokes field is excited by a quasi monochromatic single driving laser pulse, a macroscopic Stokes pulse is produced carrying with it signatures of its underlying quantum initiation in the forms of a Bose-Einstein-like probability distribution for photon number [4] and a uniform distribution of the random phase of the pulse [5]. Later it was discovered that when a Raman medium is excited by a single narrow-band laser pulse travelling in a hollow-core photonic crystal fiber (HC-PCF) containing hydrogen gas, a multi-octave optical comb spectrum, spanning the near-IR to the near-UV, can be spontaneously generated by SRS with unparalleled efficiency [6].

The present study focuses on the quantum initiation of coherence in a Raman comb spontaneously created in $H_2$-filled HC-PCF, and demonstrates for the first time an ultra-wide comb containing coherence between lines that are generated purely from quantum-initiated vacuum noise fields, using a single narrow-band pump pulse. This single-pump-laser technique contrasts with the above-mentioned Raman excitation configurations [7-9], which rely on classically exciting the medium coherence by using multiple pump lasers or ultra-short pulsed lasers. We present direct experimental confirmation of the mutual coherence of the comb spectral lines and their quantum origin. For this we show that comb lines have sufficient coherence to form correlated interference patterns. This was achieved by introducing a phase cross-correlator based on interfering independent Raman combs generated from two separate but identical Raman-scattering mediums. By comparing the fringes created simultaneously at two distinct comb frequencies, the results show a strong phase cross-correlation between any two spectral lines generated from a given Raman resonance within a single medium. These results are substantiated by a quantum theory, which, in contrast with previous works [4], shows that the quantum onset is due to spontaneously emitted light from a correlated pair of first-order Stokes (S1) and anti-Stokes (AS1) fields.

The quantum nature of the Raman comb initiation is demonstrated experimentally by the pulse-energy statistics of each spectral line from shot to shot, which exhibit the distinctive statistics of the vacuum field zero-point fluctuations. The duality of the lines' self coherence (i.e. temporal and spatial coherence within a single shot) and their super Poissonian photon statistics is attributed to the excitation condition whereby only a single vacuum spatial-temporal mode is amplified from the multimode vacuum field [10]. Furthermore, we show that the fluctuations of S1 are strongly correlated with those of the AS1. Such correlations would *not* be present in a truly thermal source; instead this shows that S1 and AS1 are part of a multimode squeezed state. This indicates the intriguing possibility of an optical comb spectrum creating attosecond pulses with nonclassical correlations (squeezing, entanglement) between the amplitudes and phases of all comb lines.

Figure 1a shows the interferometeric set-up used to measure the mutual coherence and the pulse-energy statistics of the comb components. Two 70-cm pieces of identical $H_2$-filled HC-PCF [6, 11] are excited with equal optical power from a single-frequency Nd-yttrium-aluminum-garnet laser to separately and independently generate a series of higher-order Raman spectral components. The laser emits 12 ns-long, nearly transform-limited pulses at 1064 nm with 50-Hz repetition rate. The HC-PCF used is a large-pitch single-cell Kagome fiber [6] with a 30 μm hollow core (Fig. 1b). The fiber dominantly guides a polarization-degenerate $HE_{11}$-like mode confined in the core. The two fibers are filled with hydrogen at a pressure of 15 bars. The Raman components of the generated comb from each fiber are combined on a 50/50 beamsplitter and passed through a polarizer. A grating is used to separate the Raman spectral pairs of the combined beam at the output of the interferometer, and spectral pairs are sent to a CCD camera and recorded on a single-shot basis. To ensure spatial overlap between the two beams forming the spectral line pairs, an "endlessly single-mode" PCF [12] is placed at the output of each HC-PCF to act as a spatial mode filter. The typical output spectrum from each HC-PCF is shown in Fig. 1c when the fiber is excited by ~30 mW of average pump power. The spectral comb spans from below 400 nm to above 1750 nm. The spectral components consist of Stokes and anti-Stokes lines spaced by ~125 THz for those raised from the vibrational $S_{00}(1)$ transition and by ~18 THz for those raised from the rotational $Q_{00}(1)$ [6].

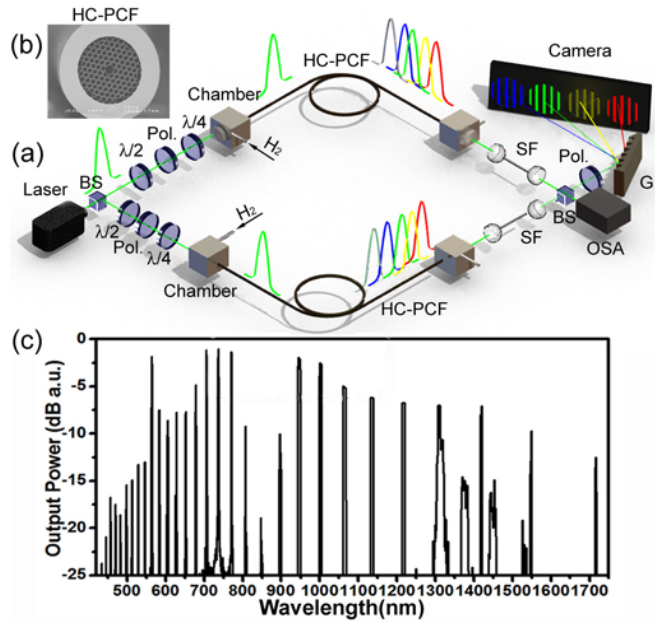

FIG.1. (a) Interferometer with $H_2$-filled HC-PCF (shown in b) in each arm. SF: spatial filter, BS: beam splitter, OSA: optical spectrum analyzer, Pol: polarizer, λ/2: half-wave plate, λ/4: quarter-wave plate, G: grating. (c) Output optical spectrum from one fiber.

Figure 2a shows single-shot profiles of the interferograms from the pump pair at 1064 nm and of four degenerate spectral pairs from the two fibers at wavelengths 740 nm, 770 nm, 1003 nm and 1134 nm. The comb-line spatial-interferograms exhibit deep interference fringes, indicating a strong "self-coherence" of each Raman line. Such interference occurs for every pair of comb components. Self coherence arises from the transiency of the Raman process [5], which in HC-PCF, due to the ultra-high Raman-gain, can occur even with a pump pulse as long as 12 ns [13]. Therefore the effects of dephasing collisions between molecules can be ignored[6]. Furthermore, because of the practically single-modedness of the fibre and the long propagation length, only one spatial mode is amplified. The net result of such a confluence in the excitation conditions is the amplification, from the multimode vacuum fields, of a single spatial-temporal mode (wave packet) that exhibits a constant phase throughout each pulse. This bears resemblance with cavity quantum electro-dynamics (cavity-QED) wherein a single molecular excitation is strongly coupled to a single cavity mode [14]. However, the difference in the present configuration is that spontaneous emission is not strongly modified as in the case of cavity-QED, but instead, and owing to the optical guiding property of the fibre and the high-gain and highly transient regime of the Raman amplification, only a single spatial-temporal mode (a "self-coherent" pulse) is *selected* from the multimode vacuum field [10].

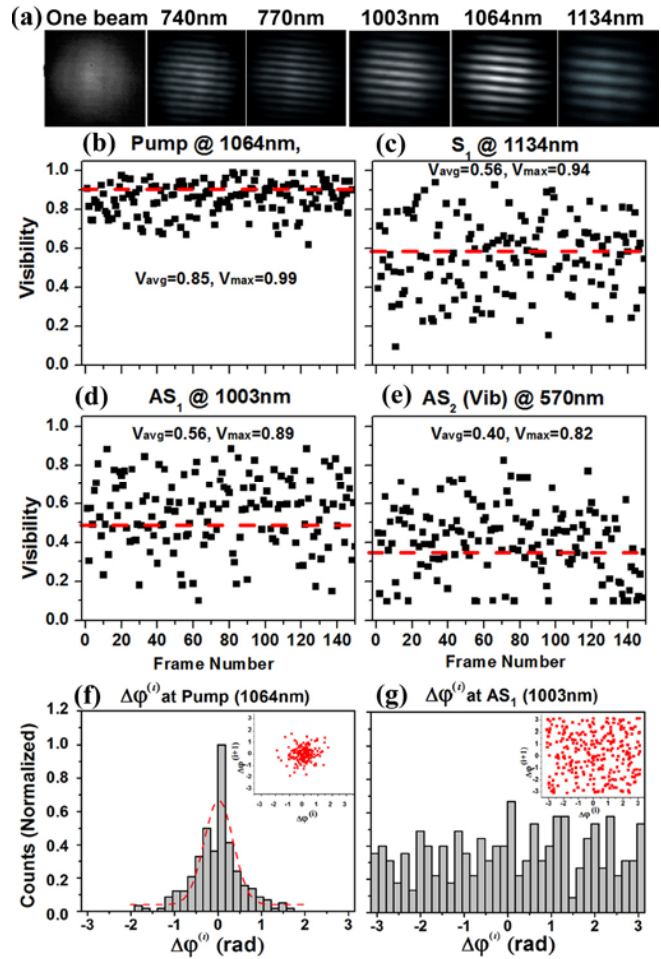

FIG. 2. (a) Mode pattern for one interferometer arm at 740 nm (far left), and (other panels) when two arm beams at the indicated wavelengths are combined. Shot-to-shot visibilities and phase distributions. (b-e) Visibilities of line-pairs from 150 different pump shots of the pump (b); two rotational (c&d) and a vibrational (e). (f,g) Phase histograms of the pump (e) and the first rotational anti-Stokes (g).

The quantum nature of the comb onset was investigated by measuring the statistical distributions of the fringe visibility and of the relative phase of each degenerate pair of comb lines from the two fibers. Figures 2b-e show the results of four representative line-pairs for 150 pump shots. For all data, we used pump pulses having the same average power (~30 mW) within a 1% range in order to reduce the effects of pump power fluctuations. The long separation between the pulses (20 ms) ensures that the spatial-temporal modes excited by each successive pump pulse are uncorrelated. Consequently, each CCD frame corresponds to a single shot of the pump laser. The visibility of each interferogram is extracted by normalizing it with the beam profile of the Raman lines for each pulse and fitting to a sine function. The fit also gives the shot-to-shot relative phase between different interferograms $\Delta\varphi^{(i)} = \varphi^{(i+1)} - \varphi^{(i)}$. The superscript (*i*) indicates the shot number or equivalently the frame number. Figures 2b & f show the shot-to-shot distributions of the visibility and phase of the transmitted pump. They both exhibit narrow distributions typical of coherent light, although the phase distribution is broadened by interferometer jitter. The measured visibility has an average of 0.85 and small standard deviation. Similarly, the phase exhibits a Gaussian distribution with a half-width-at-half-maximum of 0.78 rad and a phase portrait concentrated near the centre (inset of Fig. 2f).

In contrast, and despite exhibiting self-coherence, the higher-order Raman lines show a strongly fluctuating visibility, as illustrated by three representative components: rotational S1 at 1134nm, AS1 at 1003nm and the vibrational AS2 (Fig. 2c-e). The visibility fluctuates from near 0 to 0.94, with an average visibility of 0.40 to 0.56. The measured average visibilities are in fair agreement with the theoretical prediction for two spatially coherent, monochromatic, thermal pulses interfering, which theoretically yields a mean visibility of $\pi/4$=0.78 [4].

The phase distributions were found to be uniform for all measured Stokes and anti-Stokes lines, as illustrated by the histogram of $\Delta\varphi^{(i)}$ for AS1 at 1003nm in Fig. 2g. The observed shot-to-shot fluctuations of the visibility and phase indicate the underlying quantum nature of the S/AS lines. Moreover, we wish to emphasis that unlike in earlier experiments where the first-Stokes field is observed on its own to have thermal statistics [4], below we show that the growth of S1 is accompanied in a correlated manner with the growth of the first-order anti-Stokes (AS1) field. Such correlations would *not* be present in a thermal-like source.

Whilst the above results indicate the quantum nature of the triggering process of the comb, they don't provide information on the phase correlation (i.e. mutual coherence) between comb lines within each pulse. To that end, we observed the relationship between the fringe phases arising from a given line pair (e.g., S1 and AS1) on each shot. Correlation of these fringe phases implies correlation between lines from within a single comb. We create phase-correlation histograms by simultaneously recording interferograms of several Raman pairs within a single CCD frame. This is illustrated in Fig. 3a, which shows the interferograms for the pump P, the rotational S1, AS1 and AS2 for a given shot. Such a technique allows extracting any underlying phase relationship between any two spectral components.

In our case, as explained below, the theoretically expected phase relationship between the lines is $\varphi_n = n\varphi_{QF} + \delta_n$ [6]. Here $\delta_n$ is a deterministic phase shift that arises from the pump-laser or fiber-propagation- induced phase noise and any dynamical phase shifts from the amplification process, and $n$ is the order of the Raman lines (negative for Stokes and positive for anti-Stokes). $\varphi_{QF}$ is the resultant phase of the collective, coherent state of the molecular excitations, determined by amplification of quantum fluctuations [15], and is constant throughout a single pulse. Consequently, the phase difference for comb lines $m$ and $n$, defined as $m\varphi_n - n\varphi_m = m\delta_n - n\delta_m$, is predicted to be approximately constant.

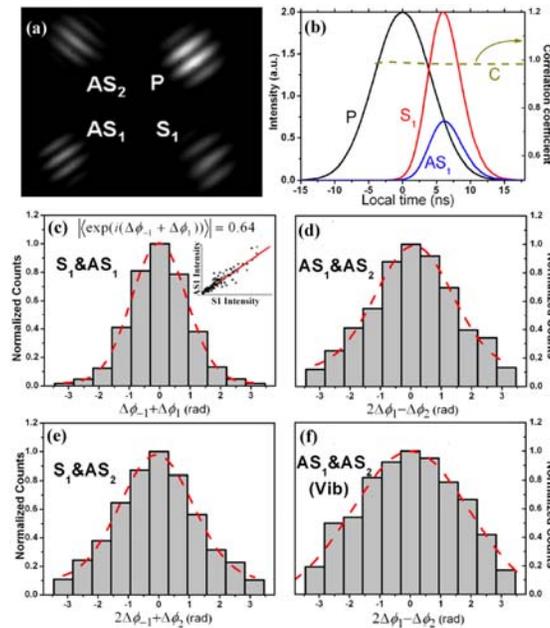

FIG. 3. Experimental and theoretical results for mutual coherence experiment. (a) interferograms of four different Raman orders in a single frame. (b) Calculated temporal profile

of mean pulse intensities of the pump (P), the first Stokes (S) and anti-Stokes (AS) as a function of local time when phase mismatch $(2\beta_p - \beta_S - \beta_{AS})L$ equals the experimental value 30. Curve C is the Stokes-anti-Stokes correlation coefficient, based on the quantum theoretical model (see text). (c) Measured histogram of phase-difference for rotational AS1 and S1, with Gaussian fit. Inset is the shot-to-shot intensity relation of the two lines. (d-f) Histograms of phase difference for (d) Rt-AS-2 and Rt-AS-1; (e) Rt-AS-2 and Rt-S-1; (f) Vb-AS-1 and Vb-AS-2.

The phase relationship mentioned above stems from the quantum propagation theory, which is based on the equations of motion for the electric-field operators $E_n^{(-)}$ of each comb line denoted by integer $n$ and the collective molecular vibration raising operator $P$ of the group of molecules located at position $z$:

$$\partial_z E_n^{(-)}(z,\tau) = -i\alpha_{2,n+1} E_{n+1}^{(-)} \exp(-i\Delta\beta_{n+1}z)P^\dagger - i\alpha_{2,n}^* E_{n-1}^{(-)} \exp(i\Delta\beta_n z)P$$

$$\partial_\tau P^\dagger(z,t) = i\sum_{n=-N}^{N} \alpha_{1,n} E_n^{(+)} E_{n-1}^{(-)} \exp(i\Delta\beta_n z) - \Gamma P^\dagger + F_\Gamma$$

Here $\alpha_{1,n}$ and $\alpha_{2,n} = 2\pi\hbar N\omega_n \alpha_{1,n}^*/c$ are complex interaction coefficients, $\Delta\beta_n = \beta_n - \beta_{n-1}$ is the difference between propagation constants of adjacent Raman lines, and $F_\Gamma$ and $\Gamma$ are the quantum Langevin operator and damping rate associated with molecular collisions. Similar equations are discussed in [6,16,17]. The initial conditions for these operators are: $\langle P^\dagger(z,0)P(z',0)\rangle = (1/AN)\delta(z-z')$, $\langle P(z,0)P^\dagger(z',0)\rangle = 0$, $\langle E_{-1,+1}^{(+)}(0,\tau')E_{-1,+1}^{(-)}(0,\tau)\rangle = (2\pi\hbar\omega_{-1,+1}/Ac)\cdot\delta(\tau-\tau')$, and $\langle E_{-1,+1}^{(-)}(0,\tau')E_{-1,+1}^{(+)}(0,\tau)\rangle = 0$. The brackets indicate expectation values, and we assumed that initially the sideband fields are in their vacuum state and all molecules are in their lowest-energy state. We consider the simplifying situation where only first-order Stokes $E_{-1}$ and anti-Stokes $E_{+1}$ fields are excited, and the pump intensity profile is unchanged throughout the interaction. This model allows us to find a full quantum description and gain insight into the comb generation process. In the high-gain, transient regime we can neglect the Langevin operator and damping. The analytic form of solutions can be found using the methods in [17,18], and take the form:

$$\begin{pmatrix} E_{-1}^{(-)}(L,\tau) \\ E_{+1}^{(+)}(L,\tau) \end{pmatrix} = E_0(\tau)\int_0^\tau d\tau' E_0(\tau') \times \begin{bmatrix} \alpha_{2,s} G_{11}(L;\tau,\tau') & \alpha_{2,s} G_{12}(L;\tau,\tau') \\ -\alpha_{2,a} G_{21}(L;\tau,\tau') & -\alpha_{2,a} G_{22}(L;\tau,\tau') \end{bmatrix}\begin{pmatrix} \alpha_{1,s} E_{-1}^{(-)}(0,\tau') \\ \alpha_{1,a} E_{+1}^{(+)}(0,\tau') \end{pmatrix},$$

$$+ E_0(\tau)\int_0^L dz' \begin{pmatrix} -i\alpha_{2,s} G_{13}(z';\tau,0) \\ i\alpha_{2,a} G_{23}(z';\tau,0) \end{pmatrix} P^\dagger(L-z',0)$$

where the Green propagators are known functions. The nonzero off-diagonal elements of the Green matrix show that the two fields are components of a squeezed state. The collective molecular excitation is also entangled in this squeezed state, making it in a sense a three-mode state having interesting quantum properties.

Figure 3b shows the calculated mean intensity profiles of the pump and rotational S1 and AS1 for the experimental conditions, but assuming only S1/AS1 are generated. They are seen to grow simultaneously. More importantly, Fig. 3b shows that the S1 and AS1 fields are predicted to have perfect phase anti-correlation, with a near-unity correlation coefficient $C$ throughout the pulse duration, which is defined by: $\left|\left\langle E_{-1}^{(-)}(L,\tau)E_{+1}^{(-)}(L,\tau)\right\rangle\right| \big/ \sqrt{\left\langle E_{-1}^{(-)}(L,\tau)E_{-1}^{(+)}(L,\tau)\right\rangle \left\langle E_{+1}^{(-)}(L,\tau)E_{+1}^{(+)}(L,\tau)\right\rangle}$. This phase anti-correlation is a result of the requirement for producing maximum Raman vibrational (or rotational) coherence in the medium, which acts as a phase-sensitive amplifier [19].

The predicted correlations between S1 and AS1 are experimentally confirmed in Fig. 3c. It shows the phase-difference histogram from 250 successive shots, with Gaussian fits centered at 0, of the first S/AS pair of the rotational Raman resonance, along with the observed shot-to-shot intensity correlation between S1 and AS1 (Fig. 3c inset). To quantify the phase correlation between the *n*-order and *m*-order Raman lines, the phase axis of the histogram is defined as $\Phi_{nm} = m\Delta\phi_n^{(i)} - n\Delta\phi_m^{(i)}$, where $\Delta\phi_n^{(i)} \equiv \Delta\varphi_n^{(i+1)} - \Delta\varphi_n^{(i)}$, and $\Delta\varphi_n^{(i)}$ is the extracted fringe phase of the *i*-th laser-shot and the *n*-order Raman line. This was evaluated by extracting interference fringe phase difference $\Delta\varphi_n^{(i)} = \varphi_{n,1}^{(i)} - \varphi_{n,2}^{(i)}$, with *1* and *2* denoting individual fibers. Consequently, under the idealization of perfect phase correlations, the above fringe phase can be rewritten as $\Delta\varphi_n^{(i)} = n\varphi_{QF,1}^{(i)} + \delta_{n,1} - (n\varphi_{QF,2}^{(i)} + \delta_{n,2}) \equiv n\Delta\varphi_{QF}^{(i)} + \Delta\delta$. The quantity $\Delta\phi_n^{(i)}$ is defined above as the phase difference of two successive shots in order to cancel the deterministic phases $\delta_n$.

The observed peaking of the distribution around zero unambiguously indicates the presence of strong phase correlation. This is quantified by the ensemble-averaged correlation coefficient $\left|\left\langle \exp(i\Phi_{nm})\right\rangle\right|$, which was found to be 0.64. Writing $I_{\pm 1} = \left\langle E_{\pm 1}^{(-)} E_{\pm 1}^{(+)}\right\rangle$, this phase correlation coefficient is directly proportional to the degree of mutual phase coherence $\left|\left\langle \exp(i\Phi_{1,-1})\right\rangle\right|$, as is seen from $C = \left|\left\langle \left|E_{-1}^{(-)}\right|\left|E_{+1}^{(-)}\right|\right\rangle\right| \left(I_{-1} I_{+1}\right)^{-1/2} \left|\left\langle e^{-i(\varphi_{-1}+\varphi_{+1})}\right\rangle\right|$, valid if we assume that the fluctuations in

intensities $\left|E_{\pm 1}^{(-)}\right|^2$ of S1 and AS1 are independent of that of their phases. The relatively smaller value 0.64 of the correlation coefficient observed for S1/AS1 (Fig. 3c) indicates that when many lines are generated, the comb dynamics are more complicated than accounted for by the idealized model, which predicts a value near 1. The observed positive correlation of S/AS intensities (Fig. 3c inset) along with anti-correlation of their phases is consistent with the behavior of a non-classical two-mode squeezed state.

Moreover, the mutual coherence is found to be present for any other spectral line pair from a given Raman resonance, as is illustrated by the phase-correlation histograms of the rotational pairs AS1/AS2 (Fig. 3d) and S1/AS2 (Fig. 3e), or the vibrational AS1/AS2 pair (Fig. 3f). These correlations follow the general relation $\varphi_n = n\varphi_{QF} + \delta_n$, mentioned above. The origin of this correlation of higher-order lines is well explained by a new self-consistent theoretical model, which will be the subject of a forthcoming publication. It shows that during the onset of S1 and AS1, the medium polarization is driven in the same manner as a phase modulator to generate the higher-order Raman sidebands. Because the S1/AS1 pair acts like a pump to generate the comb through the parametric process of phase modulation, all the higher-order components should be correlated in a manner similar to the correlation of S1/AS1.

This work has shown that an optical frequency comb spectrum spanning from the UV to the mid-IR can be spontaneously created containing a high degree of quantum coherence among comb lines, without introducing any macroscopic coherence into the system from external sources. This contrasts with other demonstrations using multiple-frequency coherent laser pulses to externally "seed" the desired coherence [3,7,8]. This opens the possibility to explore generating isolated pulses in the attosecond regime having non-classical properties such as reduced noise via quantum-state squeezing.

This work is supported by UK EPSRC grant EP/E039162/1 and US NSF grant PHY-0757818.

Correspondence and requests for materials should be addressed to F.B. (f.benabid@bath.ac.uk).